\documentclass[twocolumn,english]{revtex4-1}
\usepackage[T1]{fontenc}
\usepackage[latin9]{inputenc}
\usepackage{color}
\usepackage{bm}
\usepackage{amsmath}
\usepackage{amssymb}
\usepackage{graphicx}
\usepackage{esint}

\makeatletter

\newcommand*\LyXThinSpace{\,\hspace{0pt}}

\makeatother

\usepackage{babel}
\begin{document}

\title{Relativistic Mechanism of Chiral Magnetic Current in Weyl Semimetals
with Tilted Dispersion}

\author{Zaur Z. Alisultanov}

\affiliation{Amirkhanov Institute of Physics, Russian Academy of Sciences, Dagestan
Science Centre, Makhachkala, Russia. }

\affiliation{Dagestan State University, Makhachkala, Russia. }
\email{zaur0102@gmail.com}

\begin{abstract}
The chiral magnetic effect is a one of the exotic bulk transport properties
of the Weyl semimetals. Because of the Nielsen-Ninomiya \textquotedblleft no-go
theorem\textquotedblright , the total chiral magnetic current is absent
in the equilibrium state. One of the mechanisms for generating this
current is the chiral anomaly. This phenomenon is the anomalous nonconservation
of chiral charge for massless relativistic particles. It can be realized
by parallel magnetic and electric fields ($\sim\mathbf{E}\cdot\mathbf{\mathbf{B}}$),
and it leads to such new transport phenomenon as the negative longitudinal
magnetoresistance. Using a simple theory (we consider both linearized
and lattice model), we have shown, that in Weyl metals with tilted
dispersion another mechanism of the chiral magnetic current is possible.
It is not associated with the chiral anomaly. The new transport mechanism
is based on the relativistic effect of electric field on Landau levels.
This effect is that an electric field changes the distance between
the Landau levels, and also changes the effective velocity along magnetic
field. At presence of a tilt in the spectrum, this velocity renormalization
is differ for different Weyl points. This leads to a non-zero resulting
drift velocity. As a consequence, an electrical current arises along
the magnetic field. The induced by this mechanism the electric current
is proportional to the pseudoscalar product of the fields ($\mathbf{E}\vee\mathbf{\mathbf{B}}$)
and directed along the magnetic field, that differs it from the Hall
current ($\sim\mathbf{E}\times\mathbf{\mathbf{B}}$). At the same
time, the conductivity corresponding to this transport mechanism does
not depend on the scattering time like the Hall conductivity. Thus,
we have proposed a new anomalous transport mechanism in the Weyl semimetal,
which is not associated with the chiral anomaly.
\end{abstract}
\maketitle

\section{Introduction}

Materials with a non-trivial topology of the band structure are of
great interest for both fundamental and practical application purposes
(see, for example, \cite{key-1-1,key-2-2,key-3-3,key-4-4,key-5-5,key-6-6}).
The key property of quasi-particles in such materials is their chirality.
The main consequence of a nontrivial topology and chirality is the
topological stability (protection) of electronic states. Such stability
is based on two conditions. First, the spectrum of carriers should
be linear (Dirac fermions), since in this case chirality becomes a
good quantum number. Indeed, the massless particles retain chirality:
$\left[\mathcal{H},\gamma^{5}\right]=0$, where $\mathcal{H}=\gamma_{0}\gamma_{\mu}p^{\mu}$
is the Hamiltonian of the massless particles, $\gamma^{5}=i\gamma_{0}\gamma_{1}\gamma_{2}\gamma_{3}$
is the chirality, $\gamma_{\mu}$ are Dirac matrices. The second condition
consists in the spatial separation of states with opposite chiralities.
This condition imposes certain restrictions on the dimensions of chiral
electron states. In a 1D system, one can create zero-dimensional chiral
states localized at opposite ends. For two dimensions, opposite chiralities
can be separated to opposite edges of the plane, forming one-dimensional
edge chiral states (2D topological insulator). In a 3D system, the
opposite chiralities can be separated to opposite surfaces. The 2D
chiral states (3D topological insulator) can be created this way.
Continuing this scheme we can come to the conclusion that 3D chiral
states occur at the edges of the four-dimensional system. Formally,
such a situation has already been considered \cite{key-7-7}. However,
there is another scheme, which is more realistic and already experimentally
implemented. It consists in the fact that opposite chiralities should
be separated in some other space of parameters, for example, the momentum
space. Such materials are called 3D Weyl semimetals (WSMs) \cite{key-8-8+,key-8-8++,key-8-8,key-9-9}.
Without doping the chemical potential in such materials is localized
at the Weyl point (WP). In this paper we consider the doped WSMs with
the non-zero chemical potential, calculated from the WP. 

\begin{figure}

\includegraphics[width=6cm]{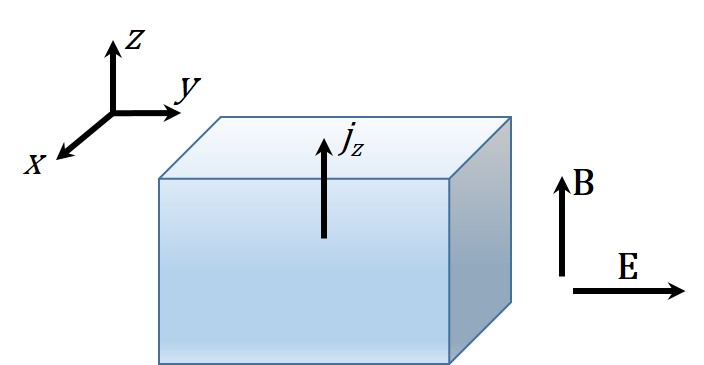}\caption{Schematic illustration of current along magnetic field. The electric
field $\mathbf{E}$ is perpendicular to the electric current $\mathbf{j}$
and to the magnetic field $\mathbf{B}$.}
\label{figure 1}

\end{figure}

The non-triviality of band structure of WSMs results in special chiral
kinetics \cite{key-10-10}. These materials have a number of exotic
transport properties, such as the anomalous Hall effect, the chiral
magnetic effect and negative magnetoresistance. In this work, we are
interested in the second of these effects. Because of the Nielsen-Ninomiya
\textquotedblleft no-go theorem\textquotedblright , the total chiral
magnetic current is absent in the equilibrium state. However, the
non-zero chiral electronic transport in the WSM can be induced by
the chiral anomaly \cite{key-10+,key-10++,key-10+++,key-11-11,key-12-12}.
This quantum phenomenon is the anomalous nonconservation of chiral
charge for Weyl particles under the parallel magnetic and electric
fields ($d\left(n_{R}-n_{L}\right)/dt\sim\mathbf{E}\cdot\mathbf{\mathbf{B}}$).
The chiral anomaly leads to such new transport phenomenon as the negative
longitudinal magnetoresistance. In this paper, using a simple theory
we have shown, that in Weyl metals with tilted dispersion another
mechanism of the chiral magnetic current is possible. It is not associated
with the chiral anomaly. The new transport mechanism is based on the
relativistic effect of electric field on Landau levels (LLs). This
effect is that an electric field changes the distance between the
LLs, and also changes the effective velocity along magnetic field.
At presence of a tilt in the spectrum, this velocity renormalization
is differ for different Weyl points. This leads to a non-zero resulting
drift velocity. As a consequence, an electrical current arises along
the magnetic field (see Fig. \ref{figure 1}). The induced by this
mechanism the electric current is proportional to the pseudoscalar
product of the fields ($\mathbf{E}\vee\mathbf{\mathbf{B}}$) and directed
along the magnetic field, that differs it from the Hall current ($\sim\mathbf{E}\times\mathbf{\mathbf{B}}$).
Thus, we have proposed a new anomalous transport mechanism in the
WSM, which is not associated with such exotic phenomena, like a chiral
anomaly. The only exotic phenomenon in the proposed mechanism consists
in the relativistic effect of the electric field on the LLs.

After introduction, the paper is organized as follows. In Sec. II
we present a general properties of tilted Weyl Hamiltonian and LLs.
We present conclusions about conditions which are imposed on the dispersion
tilt. In Sec. III we consider the relativistic effect of electric
field on the LLs and induced by such effect the new transport mechanism.
We consider both linearized (III.A) and lattice model (III.C). In
this chapter, we also give a general quasiclassical approach to the
new effect (III.B). We conclude in Sec. IV with a some evaluations
and discussion of the potential applications. 

\section{Hamiltonian and Landau levels}

\textcolor{black}{For the description of Weyl fermions with broken
Lorentz-invariance (tilted dispersion) near the i-th }WP\textcolor{black}{{}
we use a simple low-energy model \cite{key-12+}}

\begin{equation}
\mathcal{H}_{i}\left(\mathbf{p}\right)=\bm{\omega}_{i}\cdot\mathbf{p}+k_{i}\upsilon_{i}\bm{\sigma}\mathbf{p}\label{eq:1}
\end{equation}

\noindent where $k_{i}=\text{sign}\left\{ \frac{1}{2\pi}\ointop\bm{b}_{i}\left(\mathbf{p}\right)\cdot d\mathbf{S}\right\} =\pm1$
denotes the Chern number sign (chirality) of the WPs, $\bm{b}$ is
the Berry curvature, $\bm{\omega}_{i}=\left(\omega_{ix},\omega_{iy},\omega_{iz}\right)$
spectrum tilt vector, $\bm{\sigma}$ is the Pauli matrices triplet,
$\mathbf{p}$ is the quasiparticle momentum measured from WPs, $\upsilon_{i}$
is the Fermi velocity for \textcolor{black}{i-th }WP. Such a minimal
model is quite sufficient to describe the basic properties of WSMs
in a continuum approximation. The Hamiltonian (\ref{eq:1}) provides
the spectrum

\begin{equation}
\mathcal{E}_{p}^{i}=\bm{\omega}_{i}\cdot\mathbf{p}\pm\upsilon_{i}p\label{eq:2}
\end{equation}

\noindent where signs $\pm$ refer to electrons and holes respectively.
For velocity $\bm{\upsilon}=\partial\mathcal{E}_{p}/\partial\mathbf{p}$
we obtain

\begin{equation}
\bm{\upsilon}_{i}=\bm{\omega}_{i}+k_{i}\upsilon_{i}\frac{\mathbf{p}}{p}\label{eq:3}
\end{equation}

\noindent The \textquotedblleft no-go theorem\textquotedblright{}
of Nielsen-Ninomiya theorem states that in lattice gauge theory the
total Chern number of all WPs should be zero, i.e. $\sum_{i}k_{i}\upsilon_{i}=0$
(term $\bm{\omega}_{i}\cdot\mathbf{p}$ does not change the Chern
number). In particular, for two WPs this provides $\upsilon_{1,2}=\upsilon_{F}$. 

Let's find out what values of parameter $\bm{\omega}_{i}$ are allowed
to the Hamiltonian (\ref{eq:1}). For this we consider the electric
current density in the equilibrium state

\begin{equation}
\mathbf{j}_{eq}=\sum_{i}e\int f_{0}^{i}\bm{\upsilon}_{i}d^{3}\mathbf{p}\label{eq:4}
\end{equation}

\noindent where $f_{0}^{i}=\left(e^{\beta\left(\mathcal{E}_{p}^{i}-\mu\right)}+1\right)^{-1}$
is the Fermi-Dirac carrier distribution function for $i$-th WP, $\mu$
is the chemical potential. The total chiral current should be zero
in equilibrium: $\mathbf{j}_{eq}=0$. This requirement leads to the
odd the integrand 

\begin{equation}
\sum_{i}f_{0}^{i}\bm{\upsilon}_{i}\left(\mathbf{p}\right)=-\sum_{i}f_{0}^{i}\bm{\upsilon}_{i}\left(-\mathbf{p}\right)\label{eq:5}
\end{equation}
which can be transformed to 

\begin{equation}
\sum_{i}\frac{\bm{\omega}_{i}\left(x\cosh\left(\bm{\omega}_{i}\cdot\mathbf{p}\right)+1\right)-\upsilon_{F}\frac{\mathbf{p}}{p}x\sinh\left(\bm{\omega}_{i}\cdot\mathbf{p}\right)}{\left(xe^{\beta\bm{\omega}_{i}\cdot\mathbf{p}}+1\right)\left(xe^{-\beta\bm{\omega}_{i}\cdot\mathbf{p}}+1\right)}=0\label{eq:6}
\end{equation}

\noindent where we introduced $x=e^{\beta\left(\upsilon_{F}p-\mu\right)}$.
This condition can be satisfied in a variety of ways depending on
the number of WPs the system contains. The number of WPs depends on
the type of broken symmetry. If the $\mathcal{T}$ symmetry (time
reversal symmetry) is broken, the minimum number of WPs equals two.
If the $\mathcal{P}$ symmetry (space inversion symmetry) is broken,
the minimum number of points equals four. In the case with two WPs
we obtain: $\bm{\omega}_{1}=-\bm{\omega}_{2}$. In the general case,
for (\ref{eq:6}) the following condition is necessary

\noindent 
\begin{equation}
\sum_{i}\bm{\omega}_{i}=0\label{eq:7}
\end{equation}

\noindent However, it is not sufficient, because not all combinations
satisfying (\ref{eq:7}) satisfy condition (\ref{eq:6}). In the general
case, the tilt parameters are contained in the system in pairs: $\pm\bm{\omega}_{i}$,
i.e. each tilt parameter has a mirrored partner. We note that conditions
(\ref{eq:6}) does not contain the WP chirality. In case the number
of WPs is more than two, generally speaking, there are many various
options to satisfy condition (\ref{eq:6}). For example, if the number
of WPs equals four, then conditions (\ref{eq:7}) can be satisfied
with the following configurations: $\bm{\omega}_{1}=\bm{\omega}_{2}=\bm{\omega},\,\,\bm{\omega}_{3}=\bm{\omega}_{4}=-\bm{\omega}$
or $\bm{\omega}_{1}=-\bm{\omega}_{2},\,\,\bm{\omega}_{3}=-\bm{\omega}_{4}$.
etc. Each case presents its own peculiarities in relation to the effect
that is considered here. However, the essence this effect remains
the same for all cases. Therefore, below we will consider only the
case of two WPs. It is easy to generalize our analysis to the case
with $2N$ ($N\mathbb{\in Z}$) WPs. 

In magnetic field we have $\mathbf{p}\rightarrow\mathbf{p}+\frac{e}{c}\mathbf{A}$.
Let the magnetic field be directed along the Z axis, i.e. $\mathbf{\mathbf{B}}=\left(0,0,B\right)$.
We use the simplest Landau gauge: $\mathbf{A}=\left(-By,0,0\right)$.
Applying double Lorentz-boost to the Hamiltonian (\ref{eq:1}): $\mathcal{\mathcal{H}}_{i}\rightarrow e^{\frac{\sigma_{y}\lambda_{2}^{i}}{2}}e^{\frac{\sigma_{x}\lambda_{1}^{i}}{2}}\mathcal{H}_{i}e^{\frac{\sigma_{x}\lambda_{1}^{i}}{2}}e^{\frac{\sigma_{y}\lambda_{2}^{i}}{2}}$,
where $\tanh\lambda_{1}^{i}=\omega_{ix}/\upsilon_{F}$, $\tanh\lambda_{2}^{i}=\omega_{iy}/\upsilon_{F}$,
we obtain the following expression for Landau spectrum \cite{key-13-13,key-14-14,key-15-15}

\begin{equation}
\mathcal{E}_{n}^{i}=\text{sign}\left(n\right)\upsilon_{F}\sqrt{2\gamma_{i}^{3}l_{H}^{-2}\hbar^{2}n+\gamma_{i}^{2}p_{z}^{2}}+\omega_{iz}p_{z},\,\,\,\,n\neq0\label{eq:8}
\end{equation}
\begin{equation}
\mathcal{E}_{0}^{i}=\left(k_{i}\gamma_{i}\upsilon_{F}+\omega_{iz}\right)p_{z}.\label{eq:9}
\end{equation}

\noindent where $\gamma_{i}=\sqrt{1-\frac{\omega_{ix}^{2}+\omega_{iy}^{2}}{\upsilon_{F}^{2}}}$.
At presence of a magnetic field, the electric current along the $Z$
axis still does not appear. Indeed, electric current density $j_{z}$
in this case is determined as follows:

\[
j_{z}=\sum_{n,i}e\int_{-\infty}^{\infty}f_{0}^{i}\upsilon_{iz}dp_{z}.
\]

It becomes clear that the total contribution of LLs with $\upsilon_{iz}=\partial E_{n}/\partial p_{z}$
equals zero due to the fact that integrand $\sum_{i}f_{0}^{i}\upsilon_{iz}$
is odd as a result of condition (\ref{eq:6}). Certainly, the magnetic
field induces current near each WP as a result of the chiral magnetic
effect: $\mathbf{j}_{CME}=k_{i}\mathbf{B}\int\left(\frac{\mathbf{p}}{p}\cdot\bm{b}\right)f\frac{d^{3}\mathbf{p}}{\left(2\pi\right)^{3}}$.
However, the total electric current over all WPs will equal zero.
This is a consequence of the Nielsen-Ninomiya \textquotedblleft no-go
theorem\textquotedblright .

Below, for simplicity we suppose that $\omega_{iz}=0$. 

\section{Transport induced by relativistic effect of electric field}

\subsection{General consideration}

Let us now consider the effect of a perpendicular electric field on
the LLs. Let the electric field is directed along the Y axis: $\mathbf{E}=\left(0,E,0\right)$,
and the magnetic field is directed along the Z axis: $\mathbf{\mathbf{B}}=\left(0,0,B\right)$
. Then

\begin{equation}
\mathcal{\tilde{H}}_{i}\left(\mathbf{p}\right)=k_{i}\upsilon_{F}\bm{\sigma}\left(\mathbf{p}-\frac{e}{c}\mathbf{A}\right)+\bm{\omega}_{i}\cdot\left(\mathbf{p}-\frac{e}{c}\mathbf{A}\right)+eEy\label{eq:10}
\end{equation}

\begin{figure}
\includegraphics[width=8cm]{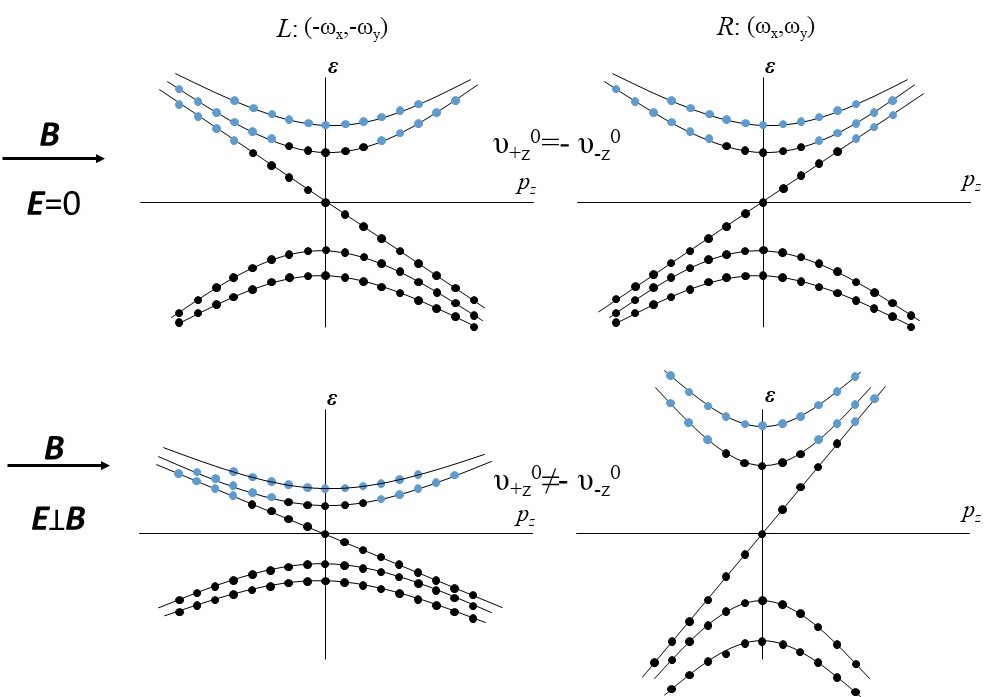}\caption{LLs (\ref{eq:8}), (\ref{eq:9}) (top part) and (\ref{eq:11}), (\ref{eq:12})
(bottom part) for pair of WPs with opposite chiralities. The tilt
parameters are given as $\bm{\omega}_{i}=\left(k_{i}\omega_{x},k_{i}\omega_{y},0\right)$.
The perpendicular electric field affects the LLs velocity in different
ways for different WPs. \label{fig:Landau levels}}
\end{figure}

\noindent Applying double Lorentz-boost $\mathcal{\mathcal{\tilde{H}}}_{i}\rightarrow e^{\frac{\sigma_{x}\tilde{\lambda}_{1}^{i}}{2}}e^{\frac{\sigma_{y}\tilde{\lambda}_{2}^{i}}{2}}\mathcal{\tilde{H}}_{i}e^{\frac{\sigma_{x}\tilde{\lambda}_{1}^{i}}{2}}e^{\frac{\sigma_{y}\tilde{\lambda}_{2}^{i}}{2}}$
with $\tanh\tilde{\lambda}_{1}^{i}=\left(\upsilon_{0}-\omega_{ix}\right)/\upsilon_{F}$,
$\tanh\tilde{\lambda}_{2}^{i}=\omega_{iy}/\upsilon_{F}$, we obtain
\cite{key-16-16,key-17-17,key-18-18}

\begin{equation}
\mathcal{\tilde{E}}{}_{n}^{i}=\text{sign}\left(n\right)\upsilon_{F}\sqrt{2\tilde{\gamma}_{i}^{3}l_{H}^{-2}\hbar^{2}n+\tilde{\gamma}_{i}^{2}p_{z}^{2}}+\upsilon_{0}p_{x},\,\,\,\,n\neq0\label{eq:11}
\end{equation}

\begin{equation}
\mathcal{\tilde{E}}{}_{0}^{i}=k_{i}\tilde{\gamma}_{i}\upsilon_{F}p_{z}+\upsilon_{0}p_{x}.\label{eq:12}
\end{equation}

\noindent where $\tilde{\gamma}_{i}=\sqrt{1-\frac{\left(\upsilon_{0}-\omega_{ix}\right)^{2}+\omega_{iy}^{2}}{\upsilon_{F}^{2}}}$,
$\bm{\upsilon}_{0}=c\frac{\left[\mathbf{E}\times\mathbf{\mathbf{B}}\right]}{B^{2}}$.
As can be seen, the electric field affects not only the cyclotron
frequency, but also the velocity along the Z-axis. This is a purely
relativistic effect. For the nonrelativistic spectrum, all the velocity
components are independent: $\upsilon_{s}=\frac{p_{s}}{m}$, $s=x,y,z$.
In the case of relativistic spectrum $\epsilon=c\sqrt{p^{2}+m^{2}c^{2}}$
each velocity component depends of all momentum components: $\upsilon_{s}=\frac{cp_{s}}{\sqrt{p^{2}+m^{2}c^{2}}}$.
Therefore, the applied electric field affects all velocity components,
although the momentum change only occur in the direction along which
the field is applied. LLs in the presence and absence of an electric
field are presented in Fig. \ref{fig:Landau levels}. The tilt parameters
in the figure are given as $\bm{\omega}_{i}=\left(k_{i}\omega_{x},k_{i}\omega_{y},0\right)$
. As can be seen from the figure, the electric field affects the LLs
velocity in different ways for different WPs since $\bm{\omega}_{1}=-\bm{\omega}_{2}$
.

The main idea of this paper can be understood already from formulas
(\ref{eq:11}),(\ref{eq:12}). The total carrier velocity contribution
from non-zero LLs still equals zero because this velocity for the
given WP is an odd function of momentum $p_{z}$. At the same time
the total velocity of zero levels from all WPs ($\upsilon_{z}^{0}=\sum_{i}k_{i}\tilde{\gamma}_{i}\upsilon_{F}$)
now does not vanish since $\tilde{\gamma}_{1}\neq\tilde{\gamma}_{2}$.
This is the result of the fact that the electric field has different
impact on tilted spectrum near different WPs. In the case with two
WPs, when $\bm{\omega}_{i}=\left(k_{i}\omega_{x},k_{i}\omega_{y},0\right)$,
we obtain

\[
\upsilon_{1z}^{0}+\upsilon_{2z}^{0}=\left[\tilde{\gamma}_{+}-\tilde{\gamma}_{-}\right]\upsilon_{F},
\]

\noindent where $\tilde{\gamma}_{\pm}=\sqrt{1-\frac{\left(\upsilon_{0}\mp\omega_{x}\right)^{2}+\omega_{y}^{2}}{\upsilon_{F}^{2}}}$.
Next, we are interested in the response of system to a weak electric
field. Then, in linear response limit for the last expression we have
the following: 

\begin{equation}
\upsilon_{1z}^{0}+\upsilon_{2z}^{0}\approx\frac{2\omega_{x}}{\gamma_{0}}\frac{\upsilon_{0}}{\upsilon_{F}}\label{eq:13}
\end{equation}

\noindent where $\gamma_{0}=\sqrt{1-\frac{\omega_{x}^{2}+\omega_{y}^{2}}{\upsilon_{F}^{2}}}$.
This expression provides the main contribution in linear approximation.
Let us consider this issue in more detail. First of all, the smallness
of the electrical field allows us to neglect the $\upsilon_{0}p_{x}$
term in the spectrum. Therefore, for the distribution function in
linear approximation we obtain

\noindent 
\begin{equation}
f^{i}=\frac{1}{e^{\beta\left(\mathcal{\tilde{E}}_{0}^{i}-\mu\right)}+1}\approx f_{0}^{i}+\delta f^{i}\frac{\omega_{x}}{\gamma_{0}}\frac{\upsilon_{0}}{\upsilon_{F}^{2}}\label{eq:14}
\end{equation}

\noindent where $\delta f^{i}=\frac{\beta\upsilon_{F}p_{z}}{4\cosh^{2}\left(\frac{1}{2}\beta\left(\mathcal{E}_{0}^{i}-\mu\right)\right)}$.
Then

\noindent 
\[
f^{1}\upsilon_{1z}^{0}+f^{2}\upsilon_{2z}^{0}\approx\left(f_{0}^{1}+f_{0}^{2}+\left(\delta f^{1}-\delta f^{2}\right)\gamma_{0}\right)\frac{\omega_{x}}{\gamma_{0}}\frac{\upsilon_{0}}{\upsilon_{F}}
\]

In the right part of the last expression we left only linear field
terms and neglected the terms that disappear after integration. Here
we should note that strictly speaking chirality conservation and absence
of scattering in WSMs are imperfect. Indeed, we can talk about chirality
conservation only when we use the low-energy approximation for Hamiltonian.
Strictly speaking, even the chiral energy level is characterized by
relaxation time $\tau_{c}$. This time can be estimated with the help
of the following expression: $\frac{1}{\tau_{c}}\sim\frac{\epsilon_{F}^{2}}{W^{2}\tau}$,
where $W$ is the parameter that characterizes the band width. However,
in this work we neglect collisions for chiral levels assuming that
the Fermi level is sufficiently close to the WP. 

So, the contribution of zero LLs to electric current density in linear
approximation is as follows: 

\begin{equation}
j_{z}^{0}=\left(\sigma_{zy}^{0}+\sigma_{zy}^{\delta}\right)E_{y}
\end{equation}

\noindent where

\noindent 
\[
\sigma_{zy}^{0}=\frac{e\omega_{x}}{\gamma_{0}}\frac{c}{\upsilon_{F}B}n_{0},
\]

\noindent 
\[
\sigma_{zy}^{\delta}=\frac{e\omega_{x}}{\gamma_{0}}\frac{c}{\upsilon_{F}B}n_{\delta},
\]

\noindent where $n_{0}=\int_{-\infty}^{\infty}\left(f_{0}^{1}+f_{0}^{2}\right)dp_{x}dp_{z}$
is the concentration of carriers on the zero LLs and $n_{\delta}=\gamma_{0}\int_{-\infty}^{\infty}\left(\delta f^{1}-\delta f^{2}\right)dp_{x}dp_{z}$.
After integration we obtain:

\noindent 
\begin{equation}
\sigma_{zy}=\sigma_{zy}^{0}+\sigma_{zy}^{\delta}=\frac{e^{2}}{\left(2\pi\hbar\right)^{2}}\frac{\omega_{x}}{\gamma_{0}^{2}}\frac{2\mu}{\upsilon_{F}^{2}}\left[\frac{\ln\left(1+e^{\beta\mu}\right)}{\beta\mu}+1\right],
\end{equation}

\noindent where we took into account that $\int dp_{x}=\frac{eB}{c}$.
We should note that this conductivity is finite in the collisionless
regime considered here. Moreover, even if the collisions are taken
into account, value $\sigma_{zy}^{0}$ does not depend on the scattering
time. This property makes the considered effect similar to the Hall
effect. We can introduce a constant similar to the Hall constant:
$R_{\omega}=\frac{E_{y}}{j_{z}^{0}H}=\frac{\upsilon_{F}\gamma_{0}}{2e\left(n_{0}+n_{\delta}\right)\omega_{x}c}$.
Such independence on the scattering sharply distinguishes this transport
mechanism from the chiral anomaly. The presence of scattering is extremely
important for observing a chiral anomaly in a condensed matter. Indeed,
due to the crystal potential periodicity, the electron dispersion
law is bounded, that will affect the continuous transfer of carriers
from one Weyl point to another. The presence of a relaxation mechanism
cuts off the effect of a periodic crystal potential. The transport
mechanism considered by us does not need an introduction of the any
relaxation mechanism. This is due to the fact that in this mechanism
the transport is determined by the pseudoscalar product of the fields
and the momentum $p_{z}$ is the integral of motion. At low temperatures
$\beta\mu\gg1$ we can evaluate this conductivity:

\begin{equation}
\sigma_{zy}\approx\frac{e^{2}}{\left(\pi\hbar\right)^{2}}\frac{\omega_{x}}{\gamma_{0}^{2}}\frac{\epsilon_{F}}{\upsilon_{F}^{2}}\label{eq:15}
\end{equation}

Now let's calculate the contribution to current from non-zero LLs.
These levels are not have chiral, so the scattering time for them
is finite. Using the $\tau$-approximation approach which is a simplest
model for collision integral, we obtain

\begin{gather}
j_{z}^{n\neq0}\propto\sum_{n,i}\int_{-\infty}^{\infty}\delta f^{i}\upsilon_{iz}^{n}\frac{dp_{x}dp_{z}}{\left(2\pi\hbar\right)^{2}}=\nonumber \\
=\frac{E_{y}}{\left(2\pi\hbar\right)^{2}}\sum_{n,i}\int dp_{x}\tau\left(\varepsilon\right)\frac{\partial}{\partial p_{y}}\int f_{0}^{i}\upsilon_{iz}^{n}dp_{z}=0\label{eq:16}
\end{gather}

The reason this contribution is absent is that the integrand expression
remains an odd function of $p_{z}$ even in the presence of an electric
field. It is important that for s-scattering even the full distribution
function remains an even function of $p_{z}$ in the presence of the
$E_{y}$ field. So, in this case formula (\ref{eq:16}) is valid not
only in linear approximation. Thus, the electric current appears only
due to zero chiral LLs. In order to completely distract ourselves
from the contribution of non-zero LLs, we consider here the case of
weakly-doped WSMs, when all electrons are located at the zero level. 

Finally, we note that term $\upsilon_{0}p_{x}$ omitted above does
not affect our calculations. If we reconstruct this term in the spectrum,
then in a linear approximation, an additional term proportional to
$\upsilon_{0}p_{x}$ appears in formula (\ref{eq:14}). However, after
integrating over $p_{x}$ this term vanishes: $\int_{-eBL_{y}/2c}^{eBL_{y}/2c}p_{x}dp_{x}=0$. 

\subsection{Quasiclassical chiral kinetic approach}

\textcolor{black}{Now we investigate our problem in a general form.
To do this, we consider the quasiclassical kinetic theory for a system
in which different WPs are characterized by the usual Weyl Hamiltonians,
but with different Fermi velocities. Hamiltonian near the i-th WP
($i=1,2$) we write in the form}

\begin{equation}
\mathcal{H}_{i}\left(\mathbf{p}\right)=k_{i}\upsilon_{i}\bm{\sigma}\mathbf{p}\label{eq:1-1}
\end{equation}

\noindent where $\upsilon_{i}$ is the Fermi velocity \textcolor{black}{in
the i-th Weyl point}. Using this Hamiltonian we write the expression
for the classic action

\[
\mathcal{S}_{i}=\int_{t_{a}}^{t_{b}}dt\left\{ \mathbf{p}\frac{d\mathbf{x}}{dt}-k_{i}\upsilon_{i}\bm{\sigma}\mathbf{p}\right\} 
\]

\noindent Using the standard method of diagonalization of the action
(see, for example, \cite{key-10-10}) and adding an electromagnetic
field, we get

\begin{equation}
\mathcal{S}_{i}=\int_{t_{a}}^{t_{b}}dt\left\{ \left(\mathbf{p}+\frac{e}{c}\mathbf{A}\right)\frac{d\mathbf{\mathbf{x}}}{dt}-e\Phi-\upsilon_{i}p-k_{i}\hbar\bm{a}_{\mathbf{p}}\cdot\frac{d\mathbf{p}}{dt}\right\} \label{eq:18}
\end{equation}

\noindent where $\Phi,\mathbf{A}$ are the scalar and vector potentials,
$\mathbf{a}_{p}=iV_{p}^{\dagger}\bm{\nabla}_{\mathbf{p}}V_{p}$ is
Berry connection and

\[
V_{p}=\frac{1}{\sqrt{2p\left(p-p_{z}\right)}}\left(\begin{array}{c}
p_{x}-ip_{y}\\
p-p_{z}
\end{array}\begin{array}{c}
-p+p_{z}\\
p_{x}+ip_{y}
\end{array}\right).
\]

In equilibrium state, as we saw above, $\upsilon_{i}=\upsilon_{F}$.
Now we use this parameter in general form, assuming that the Fermi
velocities of different points differ in magnitude for some reason
(of course, this state is not equilibrium). The Eq.(\ref{eq:18})
gives the following equations of motion (we use the Hamilton formalism)

\[
\frac{d\mathbf{p}}{dt}=\frac{e}{c}\left[\frac{d\mathbf{x}}{dt}\times\mathbf{B}\right]+e\mathbf{E}
\]

\[
\frac{d\mathbf{x}}{dt}=\upsilon_{i}\frac{\mathbf{p}}{p}+k_{i}\hbar\left[\frac{d\mathbf{p}}{dt}\times\bm{b}\right]
\]

\noindent where $\bm{b}=\bm{\nabla}_{\mathbf{p}}\times\mathbf{a}_{p}=\mathbf{p}/2p^{3}$
is the Berry curvature. Using these equations you can get the \textcolor{black}{quasiclassical}
kinetic equation and current density

\begin{gather}
\mathbf{j}_{i}=e\upsilon_{i}\int_{\mathbf{p}}\frac{\mathbf{p}}{p}f+k_{i}\hbar e^{2}\left[\mathbf{E}\times\int_{\mathbf{p}}f\bm{b}\right]+\nonumber \\
+k_{i}\upsilon_{i}\frac{e^{2}}{c}\hbar\mathbf{B}\int_{\mathbf{p}}f\left(\frac{\mathbf{p}}{p}\cdot\bm{b}\right)
\end{gather}

\noindent The first term of right part describes the usual non-chiral
current. The contribution from nonzero LLs, which we calculated in
Eq.(\ref{eq:16}) refers to this term. The second term describes the
anomalous quantum Hall effect. Note that this effect is determined
by the Berry curvature and the distance between the WPs. The relativistic
effect of the electric field, discussed here, has no affect this term.
The third term describes the chiral-magnetic effect

\begin{equation}
\mathbf{j}_{CME}^{i}=k_{i}\upsilon_{i}\frac{e^{2}}{c}\hbar\mathbf{B}\int_{\mathbf{p}}f\left(\frac{\mathbf{p}}{p}\cdot\bm{b}\right)
\end{equation}

The chiral magnetic current in this form depends both on the sign
($k_{i}$) and on the magnitude ($\upsilon_{i}$) of the velocity.
This expression can be reduced to the more simple and clear form at
zero temperature. Taking into account that

\begin{equation}
\int_{\mathbf{p}}f\left(\frac{\mathbf{p}}{p}\cdot\bm{b}\right)=\upsilon_{i}^{2}\int_{0}^{\mu_{i}}\frac{\rho\left(\epsilon\right)d\epsilon}{2\epsilon^{2}}=\frac{\mu_{i}}{4\pi^{2}\hbar^{3}\upsilon_{i}},
\end{equation}

\noindent we get that the chiral current is determined by the chemical
potentials of WPs

\begin{equation}
j_{CME}=\frac{e^{2}}{4\pi^{2}\hbar^{2}c}\sum_{i}k_{i}\mu_{i}B.\label{eq:24-1}
\end{equation}

This shows that the total chiral current is nonzero only if the chemical
potentials of different WPs are different: $\mu_{1}\neq\mu_{2}$.
This can be achieved, for example, by chiral anomaly. In this case

\[
\mu_{i}=\mu_{0}+k_{i}\upsilon_{i}eE\tau,
\]

\noindent where $\tau$ is the transport relaxation time. Then

\begin{equation}
j_{CAE}=\frac{e^{3}}{4\pi^{2}\hbar^{2}c}\left(\upsilon_{1}+\upsilon_{2}\right)\tau EB\label{eq:23}
\end{equation}

\noindent As we can see, this current depends on the Fermi velocity
at both WPs. If $\upsilon_{1}=\upsilon_{2}=\upsilon_{F}$ we get the
well-known formula for anomaly-induced current.

Let us now apply Eq.(\ref{eq:24-1}) to describe the relativistic
mechanism of chiral transport. As shown above, the transverse electric
field changes the velocity of the zero LL. I.e. the electric field
is an external perturbation that leads to inequality of velocities
$\upsilon_{1}\neq\upsilon_{2}$. It is easy to show that this leads
to the inequality of the chemical potentials at different WPs, i.e.
$\mu_{1}\neq\mu_{2}$. In other words, the equality of chemical potentials
in different Weyl cones can be violated not only by the imbalance
of the chiral charge (chiral anomaly), but also by changing the angles
of these cones (Fermi velocities). Indeed, in the absence of an electric
field we have $\upsilon_{1}=\upsilon_{2}=\upsilon_{F}$. Then the
carrier concentration near each of the Weyl points is defined as follow

\[
n=\int_{0}^{p_{0}}\frac{4\pi p^{2}dp}{\left(2\pi\hbar\right)^{3}}=\frac{1}{2\pi^{2}\hbar^{3}\upsilon_{F}^{3}}\frac{\mu_{0}^{3}}{3},
\]

\noindent or

\[
\mu_{0}=\upsilon_{F}\hbar\left(6\pi^{2}n\right)^{1/3}.
\]

\noindent The electric field leads to a violation of the equality
of velocities, i.e. $\upsilon_{1}\neq\upsilon_{2}$. Consequently,

\begin{gather*}
\mu_{1}=\upsilon_{1}\hbar\left(6\pi^{2}n\right)^{1/3}=\frac{\mu_{0}}{\upsilon_{F}}\upsilon_{1}\\
\mu_{2}=\upsilon_{2}\hbar\left(6\pi^{2}n\right)^{1/3}=\frac{\mu_{0}}{\upsilon_{F}}\upsilon_{2}
\end{gather*}

\noindent Then

\begin{equation}
j_{CME}=\frac{e^{2}}{4\pi^{2}\hbar^{2}}\frac{\mu_{0}}{\upsilon_{F}}B\left(\upsilon_{1}-\upsilon_{2}\right)\label{eq:24}
\end{equation}

\noindent This is the semiclassical explanation of the effect discussed
above. Using this formula and taking into account (\ref{eq:13}) we
can easily obtain the formula (\ref{eq:15}). Of course, it should
be noted that the current (\ref{eq:24}) disappears in the absence
of doping, while the anomaly-induced current does not depend on the
doping.

Once again we note that an important difference between relativistic
mechanism and chiral anomaly is that first of them does not need an
introduction of the relaxation time, while it is vital for the second.

Note that in a more rigorous model we need to take into account that
the two WPs are connected to each other. Indeed, the WPs are different
points of a single continuous Brillouin zone. In order to take this
into account, we need to consider the lattice model. The calculations
for the lattice model in the simplest form are given in the next chapter.

\subsection{Calculations for lattice model}

Finally, we briefly generalize the above results for the lattice model.
We use the simplest model of the Weyl semimetal with a broken $\mathcal{T}$
symmetry, namely, the Burkov-Balent model. This model based on a magnetically
doped TI-NI (TI is topological insulator and NI is normal insulator)
multilayer heterostructure. This Hamiltonian has the form \cite{key-25+}

\noindent 
\begin{equation}
\mathcal{\mathcal{H}}_{BB}=\upsilon_{F}\tau^{z}\left[\hat{z}\times\bm{\sigma}\right]\cdot\mathbf{p}_{\perp}+m\sigma_{z}+\hat{\Delta}\left(k_{z}\right)\label{eq:25}
\end{equation}

\noindent where $\hat{\Delta}\left(k_{z}\right)=\Delta_{S}\tau^{x}+\frac{1}{2}\left(\Delta_{D}\tau^{+}e^{ik_{z}d}+\text{h.c.}\right)$,
$\mathbf{p}_{\perp}=\left(p_{x},p_{y}\right)$ is the momentum in
the 2D surface Brillouin zone, $\bm{\tau}=\left(\tau_{x},\tau_{y},\tau_{z}\right)$
are Pauli matrices, acting on the which surface pseudospin degree
of freedom, $m$ describes exchange spin splitting of the surface
states, $\Delta_{S}$ and $\Delta_{D}$ describe tunneling between
top and bottom surfaces within the same TI layer ($\Delta_{S}$),
and between top and bottom surfaces of neighboring TI layers ($\Delta_{D}$).
Here we consider the case when $\Delta_{D}$ and $\Delta_{S}$ have
opposite signs.

\begin{figure}
\includegraphics[width=4cm]{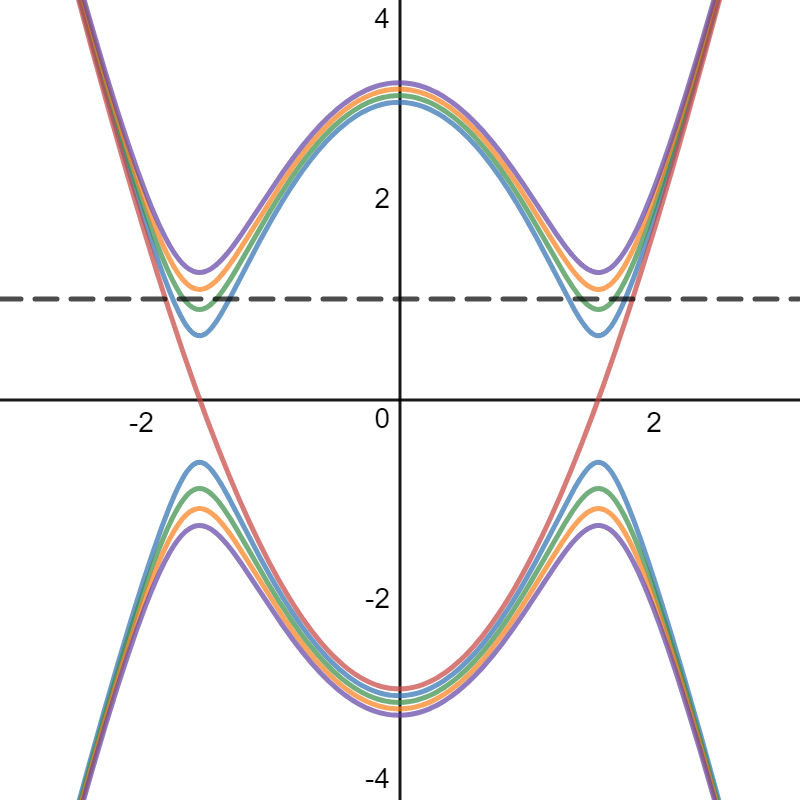}
\includegraphics[width=4cm]{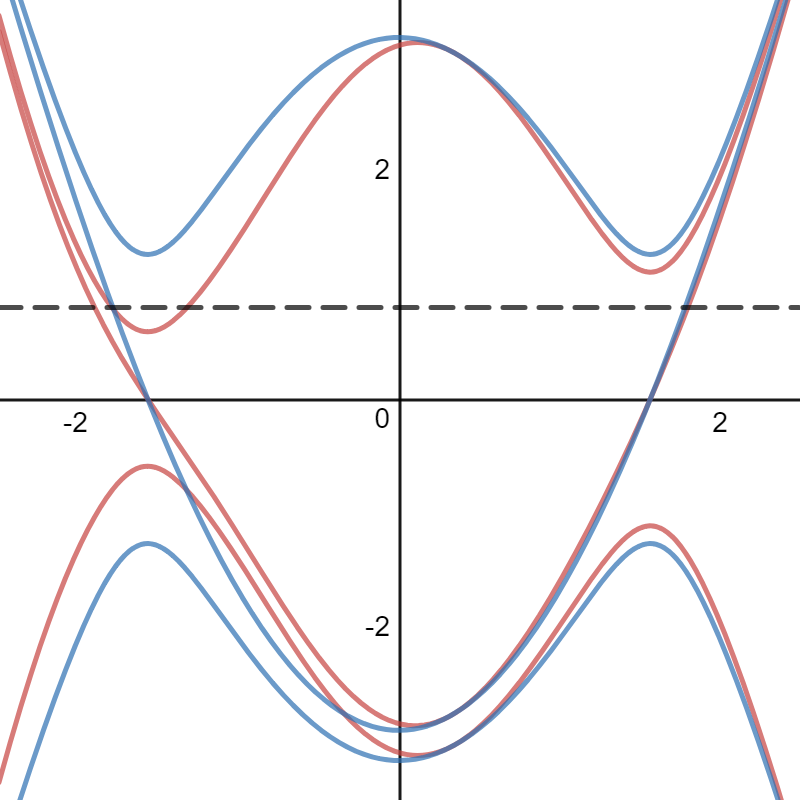}\caption{Qualitative picture of LLs in the lattice model. The left side shows
LLs in the absence of an electric field along the Y axis. The right
shows LLs in the presence of an electric field. Due to the presence
of tilt in the spectrum, the electric field has a different effect
on LLs near different WPs. In other words, the symmetric levels become
asymmetrical. The degree of such asymmetry of the zero LL determines
the magnitude of the resulting chiral current. \label{fig:Qualitative-picture-of}}
\end{figure}

\noindent Now we generalize this model by adding a tilt term in simplest
form $\bm{\omega}_{i}=\left(\omega_{x}\left(k_{z}\right),0,0\right)$.
A simple analysis shows that the parameter $\omega_{x}$ in the lattice
model (\ref{eq:25}) must be a periodic function of $k_{z}$. This
function can be defined as $\omega_{x}\left(k_{z}\right)=\omega_{x}\sin\left(\frac{\pi k_{z}}{2k_{0}}\right)$.
At Weyl points $\pm k_{0}$ we get $\omega_{x}\left(k_{z}\right)=\pm\omega_{x}$,
and in between these points the function $\omega_{x}\left(k_{z}\right)$
changes smoothly. Then for the tilted Hamiltonian we can write

\noindent 
\begin{gather}
\mathcal{\mathcal{\tilde{H}}}_{BB}=\upsilon_{F}\tau^{z}\left[\hat{z}\times\bm{\sigma}\right]\cdot\left(\mathbf{p}+\frac{e}{c}\mathbf{A}\right)+m\sigma_{z}+\nonumber \\
+\omega_{x}\sin\left(\frac{\pi k_{z}}{2k_{0}}\right)\left(p_{x}+\frac{e}{c}A_{x}\right)+\hat{\Delta}\left(k_{z}\right)+eEy\label{eq:26}
\end{gather}

\noindent where we have included electric and magnetic fields in the
same geometry as above in text, $\mathbf{A}=\left(-By,0,0\right)$.
Applying now a non-uniform Lorentz-boost

\noindent 
\[
\mathcal{\mathcal{\tilde{H}}}_{BB}\rightarrow e^{\tau^{z}\sigma_{y}\frac{\tilde{\lambda}_{1}}{2}}\mathcal{\tilde{H}}_{BB}e^{\tau^{z}\sigma_{y}\frac{\tilde{\lambda}_{1}}{2}}
\]

\noindent with $\tanh\tilde{\lambda}_{1}=\left(\omega_{x}\sin\left(\frac{\pi k_{z}}{2k_{0}}\right)-\upsilon_{0}\right)/\upsilon_{F}=\tilde{\beta}$,
we reduce the eigenvalues problem of Hamiltonian (\ref{eq:26}) to
eigenvalues problem of such Hamiltonian without electric field and
with effective magnetic field $B\sqrt{1-\tilde{\beta}^{2}}$. The
eigenvalue problem for the Hamiltonian (\ref{eq:25}) in a quantizing
magnetic field was solved in the paper \cite{key-29}. After the diagonalization
of the operator $\hat{\Delta}\left(k_{z}\right)$, the Hamiltonian
can be represented in block form. This leads to four branches of the
spectrum. We are interested only in those branches that contain WPs.
Considering only such solutions, for the Hamiltonian spectrum (\ref{eq:26})
we finally get

\begin{equation}
\mathcal{\tilde{E}}{}_{n}=\pm\upsilon_{F}\sqrt{2\tilde{\gamma}_{L}^{3}l_{H}^{-2}\hbar^{2}n+\tilde{\gamma}_{L}^{2}\left(m-\Delta\left(p_{z}\right)\right)^{2}}+\upsilon_{0}p_{x},\label{eq:11-1}
\end{equation}

\begin{equation}
\mathcal{\tilde{E}}{}_{0}=\tilde{\gamma}_{L}\left(m-\Delta\left(p_{z}\right)\right)+\upsilon_{0}p_{x}.\label{eq:12-1}
\end{equation}

\noindent where $\tilde{\gamma}_{L}=\sqrt{1-\frac{\left(\omega_{x}\sin\left(\frac{\pi k_{z}}{2k_{0}}\right)-\upsilon_{0}\right)^{2}}{\upsilon_{F}^{2}}}$,
$m<\Delta\left(\frac{\pi}{d}\right)$, $m\neq0$, $\Delta\left(k_{z}\right)=\sqrt{\Delta_{S}^{2}+\Delta_{D}^{2}+2\Delta_{D}\Delta_{S}\cos\left(k_{z}d\right)}$.
The form of the spectrum at $\upsilon_{0}=0$ is shown in Fig. \ref{fig:Qualitative-picture-of}a.
Due to the symmetry of the zones, the total chiral magnetic current
is zero. In Fig. \ref{fig:Qualitative-picture-of}b shows the zero
and first LLs in the absence and presence of an electric field. As
we can see, the levels become asymmetrical when the electric field
is turned on. The degree of such asymmetry of the Landau zero level
determines the resulting chiral current. Note that the application
of the electric field does not change the distance between the WPs.
This means that the relativistic influence of the electric field discussed
here has no effect on the anomalous quantum Hall effect. 

As we can see from Fig.\ref{fig:Qualitative-picture-of}b, the non-zero
LLs are still quite symmetrical with respect to the Weyl point for
small values of the chemical potential, as well as for small values
of $\upsilon_{0}$ (we are interested in weak electric field). This
means that the contribution from these levels to the total current
is negligible. In addition, these levels are not chiral and are sensitive
to scattering. Therefore, we consider only the contribution from the
zero LLs.

\begin{gather}
j_{z}^{0}=\frac{e^{2}}{4\pi^{2}\hbar^{2}}\frac{B}{c}\int_{-\hbar\pi/d}^{\hbar\pi/d}f\upsilon_{z}^{0}dp_{z}=\nonumber \\
=\frac{e^{2}}{4\pi^{2}\hbar^{2}}\frac{B}{c}\frac{1}{\beta}\ln\frac{e^{-\beta\left(\Gamma_{+}-\mu\right)}+1}{e^{-\beta\left(\Gamma_{-}-\mu\right)}+1}\label{eq:31}
\end{gather}

\noindent where $\Gamma_{\pm}=\sqrt{1-\frac{\left(\omega_{x}\sin\left(\frac{\pi^{2}}{2k_{0}d}\right)\mp\upsilon_{0}\right)^{2}}{\upsilon_{F}^{2}}}\left(m-\Delta\left(\frac{\pi}{d}\right)\right)$.
The current vanishes at $\upsilon_{0}=0$. In the linear response
approximation, we obtain the following expression for the conductivity

\begin{gather*}
\sigma_{zy}=-\frac{e^{2}}{\pi^{2}\hbar^{2}}\frac{\omega_{x}\sin\left(\frac{\pi^{2}}{2k_{0}d}\right)}{\upsilon_{F}^{2}\Gamma_{0}}\frac{\left(\Delta\left(\frac{\pi}{d}\right)-m\right)^{2}}{e^{\beta\left(\Gamma_{0}-\mu\right)}+1}
\end{gather*}

\noindent where $\Gamma_{0}=\sqrt{1-\frac{\omega_{x}^{2}\sin^{2}\left(\frac{\pi^{2}}{2k_{0}d}\right)}{\upsilon_{F}^{2}}}\left(m-\Delta\left(\frac{\pi}{d}\right)\right)$. 

\section{Conclusion}

Thus, in WSMs with a tilted spectrum apart from the well-known mechanism
of transport induction due to the chiral anomaly ($\sim\mathbf{E}\cdot\mathbf{\mathbf{B}}$)
there is also another mechanism caused by the renormalization of LLs
and proportional to the pseudoscalar product of magnetic and electric
fields ($\sim\mathbf{E}\vee\mathbf{\mathbf{B}}$). At an arbitrary
angle between the electric and magnetic fields (when both field-configurations
$\mathbf{E}\cdot\mathbf{\mathbf{B}}$ and $\mathbf{E}\vee\mathbf{\mathbf{B}}$
are non-zero), the total current should consist of two contributions 

\begin{equation}
j_{\mathbf{B}}=\sigma_{RE}\frac{\mathbf{E}\vee\mathbf{B}}{B}+\sigma_{CAE}\frac{\mathbf{E}\cdot\mathbf{B}}{B},\label{eq:32}
\end{equation}

\noindent where ``$\mathbf{\mathbf{B}}$'' index means that the
electric current is directed along the magnetic field, $\sigma_{RE}$
is the conductivity induced by the previously described effect. ``$RE$''
index stands for \textquotedblleft Relativistic Effect\textquotedblright ,
because the corresponding part of the electric current results from
the relativistic effect of the electric field\textquoteright s impact
on LLs. Note that the first term in (\ref{eq:32}) changes its sign
when the vector $\mathbf{E}$ is reflected with respect to the direction
of the vector $\mathbf{\mathbf{B}}$\textcolor{black}{{} ($\alpha\rightarrow2\pi-\alpha$,
where $\alpha$ is the angle between the $\mathbf{E}$ and $\mathbf{\mathbf{B}}$
vectors) and the sign of the second term remains the same. At the
same time, when the vector $\mathbf{E}$ is reflected with respect
to perpendicular to the magnetic field axis ($\alpha\rightarrow\pi-\alpha$),
the sign of second term changes, and the first remains the same.}

Let's make some estimations of the predicted conductivity. We will
consider the case with zero temperature. We suppose that $\epsilon_{F}=100\text{meV}$,
$\upsilon_{F}=10^{8}\text{cm/sec}$, $\omega_{x}=\omega_{y}=0.5\upsilon_{F}$.
We should note that with these parameters the energy of the first
Landau level exceeds the Fermi level in fields $B>1\text{T}$. In
this case we can take into account only the zero LL and use formula
(11): $\sigma_{zy}\simeq10^{13}\text{sec}^{-1}$. Using the formula
(\ref{eq:13}) we can estimate the carrier mobility through the considered
transport channel. For the parameter values we use here the estimations
provide value $\mu_{zy}\simeq10^{4}\frac{\text{cm}^{2}}{\text{V}\cdot\text{sec}}$
at $B=1\text{T}$. Note that conductivity $\sigma_{zy}$ decreases
with increasing magnetic field intensity. This can be explained that
the difference between velocities of zero LLs for different WPs decreases
as well. 

The effect described above results from chiral zero LLs. This is precisely
the reason we called it the chiral current. For example, in a Dirac
semimetal this effect should be non-existent. 

We should note that such current will not appear with respect to gauge
pseudo-fields induced by deformations \cite{key-19,key-20,key-21,key-22,key-23}.
Indeed, pseudo-fields appear with different signs in different WPs.
Then $\upsilon_{0}^{+}=-\upsilon_{0}^{-}$. In this case LLs for both
WPs are renormalized in an identical manner. In addition, the considered
effect will be absent in certain special cases of systems with $4N$
WPs. The following configuration is an example for such case: $\bm{\omega}_{+}^{(1)}=\bm{\omega}_{-}^{(1)}=\left(\omega_{x},\omega_{y},0\right),\,\,\bm{\omega}_{+}^{(2)}=\bm{\omega}_{-}^{(2)}=\left(-\omega_{x},-\omega_{y},0\right)$,
where superscripts (1,2) denominate the Weyl pair numbers. 

Finally, the fact that the proposed transport mechanism is directly
related to the tilt in the spectrum gives potential application possibilities.
In particular, the study of this transport can become an effective
tool for identify and study of the tilt in the spectrum. By changing
the directions of the electric and magnetic fields, one can get quite
rich information about the tilt parameters.We should note that considering
a stricter theory of the describe effect is of certain interest. The
stricter theory should account for tilt along the $Z$ axis as well.
The latter may result in a non-zero contribution from non-zero LLs.
Moreover, the Shubnikov-de Haas oscillations should be accounted for.
All these effect are of paramount interest, but the essence of the
effect described in this work evidently will not change at the qualitative
level. 

I thank Dmitry Kharzeev and Alexey Soluyanov for the useful discussion
of the work. This work was supported by grants RFBR 18-02-01022a,
18-32-00205 mol(a), 19-02-01000 A.

\end{document}